\documentclass[twocolumn,tighten]{aastex63}
\hypersetup{linkcolor=red,citecolor=blue,filecolor=cyan,urlcolor=magenta}
\usepackage{amsmath}
\usepackage{natbib}
\usepackage{enumitem}

\newcommand{\hdone}{\object[HD 19445]{HD~19445}}
\newcommand{\hdeight}{\object[HD 84937]{HD~84937}}
\newcommand{\hdonezero}{\object[HD 108317]{HD~108317}}
\newcommand{\hdonetwo}{\object[HD 128279]{HD~128279}}
\newcommand{\hdonefour}{\object[HD 140283]{HD~140283}}
\newcommand{\hdonesix}{\object[HD 160617]{HD~160617}}
\newcommand{\hdtwo}{\object[HD 222925]{HD~222925}}
\newcommand{\bd}{\object[BD+17 3248]{BD~$+$17$^{\circ}$3248}}

\newcommand{\loggf}{\mbox{$\log(gf)$}}
\newcommand{\kmsec}{\mbox{km~s$^{\rm -1}$}}
\newcommand{\logg}{\mbox{log~{\it g}}}
\newcommand{\msun}{\mbox{$M_{\odot}$}}
\newcommand{\teff}{\mbox{$T_{\rm eff}$}}
\newcommand{\vt}{\mbox{$v_{\rm t}$}}
\newcommand{\rpro}{\mbox{{\it r}-process}}
\newcommand{\spro}{\mbox{{\it s}-process}}
\newcommand{\ipro}{\mbox{{\it i}-process}}

\newcommand{\logeps}[1]{$\log\varepsilon$(#1)}

\shorttitle{The First and Second R-process Peaks}
\shortauthors{Roederer et al.}
\accepted{for publication in the Astrophysical Journal (ApJ 936, 84 (2022))}

\begin{document}

\title{%
The $R$-Process Alliance:\
Abundance Universality among Some Elements \\
at and between the First and Second $R$-Process Peaks%
\footnote{%
Based on archival observations made with the NASA/ESA 
Hubble Space Telescope
(program GO-7348,
program GO-8342,
program GO-9455,
program GO-14161, and
program GO-14672)
and observations collected from the 
Keck Observatory Archive
(program U25H).}}

\author{Ian U.\ Roederer}
\affiliation{%
Department of Astronomy, University of Michigan,
1085 S.\ University Ave., Ann Arbor, MI 48109, USA}
\affiliation{%
Joint Institute for Nuclear Astrophysics -- Center for the
Evolution of the Elements (JINA-CEE), USA}
\email{Email:\ iur@umich.edu}

\author{John J.\ Cowan}
\affiliation{Homer L.\ Dodge Department of Physics and Astronomy, 
University of Oklahoma, Norman, OK 73019, USA}

\author{Marco Pignatari}
\affiliation{%
Konkoly Observatory, Research Centre for Astronomy and Earth Sciences, 
E\"otv\"os Lor\'and Research Network (ELKH), 
MTA Centre of Excellence, 
Konkoly Thege Mikl\'{o}s \'{u}t 15-17, 
H-1121 Budapest, Hungary}
\affiliation{%
E.A. Milne Centre for Astrophysics, Dept of Physics \& Mathematics, 
University of Hull, HU6 7RX, United Kingdom}
\affiliation{%
NuGrid Collaboration, \url{http://nugridstars.org}}
\affiliation{%
Joint Institute for Nuclear Astrophysics -- Center for the
Evolution of the Elements (JINA-CEE), USA}

\author{Timothy C.\ Beers}
\affiliation{%
Department of Physics and Astronomy, University of Notre Dame, 
Notre Dame, IN 46556, USA}
\affiliation{%
Joint Institute for Nuclear Astrophysics -- Center for the
Evolution of the Elements (JINA-CEE), USA}

\author{Elizabeth A.\ Den Hartog}
\affiliation{%
Department of Physics, University of Wisconsin-Madison,
Madison, WI 53706, USA}

\author{Rana Ezzeddine}
\affiliation{%
Department of Astronomy, University of Florida, Bryant Space Science Center,  
Gainesville, FL 32611, USA}
\affiliation{%
Joint Institute for Nuclear Astrophysics -- Center for the
Evolution of the Elements (JINA-CEE), USA}

\author{Anna Frebel}
\affiliation{%
Department of Physics and Kavli Institute for Astrophysics and Space Research, 
Massachusetts Institute of Technology, 
Cambridge, MA 02139, USA}
\affiliation{%
Joint Institute for Nuclear Astrophysics -- Center for the
Evolution of the Elements (JINA-CEE), USA}

\author{Terese T.\ Hansen}
\affiliation{%
Department of Astronomy, 
Stockholm University, 
SE-106 91 Stockholm, Sweden}

\author{Erika M.\ Holmbeck}
\affiliation{%
Carnegie Observatories,
Pasadena, CA 91101, USA}
\affiliation{%
Joint Institute for Nuclear Astrophysics -- Center for the
Evolution of the Elements (JINA-CEE), USA}

\author{Matthew R.\ Mumpower}
\affiliation{%
Theoretical Division, Los Alamos National Laboratory,
Los Alamos, New Mexico 87545, USA}
\affiliation{%
Center for Theoretical Astrophysics, Los Alamos National Laboratory, 
Los Alamos, New Mexico 87545, USA}
\affiliation{%
Joint Institute for Nuclear Astrophysics -- Center for the
Evolution of the Elements (JINA-CEE), USA}

\author{Vinicius M.\ Placco}
\affiliation{%
NSF's NOIRLab, 
Tucson, AZ 85719, USA}

\author{Charli M.\ Sakari}
\affiliation{%
Department of Physics and Astronomy, San Francisco State University,
San Francisco, CA 94132, USA}

\author{Rebecca Surman}
\affiliation{%
Department of Physics, University of Notre Dame, 
Notre Dame, IN 46556, USA}
\affiliation{%
Joint Institute for Nuclear Astrophysics -- Center for the
Evolution of the Elements (JINA-CEE), USA}

\author{Nicole Vassh}
\affiliation{%
TRIUMF,
Vancouver, BC V6T 2A3, Canada}

\begin{abstract}

We present new observational benchmarks
of rapid neutron-capture process (\rpro)
nucleosynthesis for elements
at and between the first ($A \sim 80$) and
second ($A \sim 130$) peaks.
Our analysis is based on
archival ultraviolet and optical spectroscopy
of eight metal-poor stars
with Se ($Z = 34$) or Te ($Z = 52$) detections,
whose \rpro\ enhancement varies by more than a factor of 30
($-0.22 \leq$ [Eu/Fe] $\leq +1.32$).
We calculate ratios among the abundances of
Se,
Sr through Mo ($38 \leq Z \leq 42$), and
Te.
These benchmarks may offer a new empirical alternative 
to the predicted solar system \rpro\ residual pattern.
The Te abundances in these stars correlate more closely with 
the lighter \rpro\ elements
than the heavier ones, 
contradicting and superseding previous findings.
The small star-to-star dispersion among the abundances of
Se, Sr, Y, Zr, Nb, Mo, and Te
($\leq$~0.13~dex, or 26\%) 
matches that
observed among the abundances of 
the lanthanides and third \rpro-peak elements.
The concept of \rpro\ universality that is recognized among the
lanthanide and third-peak elements in \rpro-enhanced stars
may also apply to Se, Sr, Y, Zr, Nb, Mo, and Te,
provided the overall abundances of the lighter \rpro\ elements
are scaled independently of the heavier ones.
The abundance behavior of the elements Ru through Sn
($44 \leq Z \leq 50$) requires further study.
Our results suggest that
at least one relatively common source in the early Universe
produced a consistent abundance pattern
among some elements spanning
the first and second \rpro\ peaks.

%%%%% PLAIN-TEXT ABSTRACT
%We present new observational benchmarks of rapid neutron-capture process (r-process) nucleosynthesis for elements at and between the first (A ~ 80) and second (A ~ 130) peaks. Our analysis is based on archival ultraviolet and optical spectroscopy of eight metal-poor stars with Se (Z = 34) or Te (Z = 52) detections, whose r-process enhancement varies by more than a factor of 30 (-0.22 <= [Eu/Fe] <= +1.32). We calculate ratios among the abundances of Se, Sr through Mo (38 <= Z <= 42), and Te. These benchmarks may offer a new empirical alternative to the predicted solar system r-process residual pattern. The Te abundances in these stars correlate more closely with the lighter r-process elements than the heavier ones, contradicting and superseding previous findings. The small star-to-star dispersion among the abundances of Se, Sr, Y, Zr, Nb, Mo, and Te (<= 0.13 dex, or 26%) matches that observed among the abundances of the lanthanides and third r-process-peak elements. The concept of r-process universality that is recognized among the lanthanide and third-peak elements in r-process-enhanced stars may also apply to Se, Sr, Y, Zr, Nb, Mo, and Te, provided the overall abundances of the lighter r-process elements are scaled independently of the heavier ones. The abundance behavior of the elements Ru through Sn (44 <= Z <= 50) requires further study. Our results suggest that at least one relatively common source in the early Universe produced a consistent abundance pattern among some elements spanning the first and second r-process peaks.
%%%%%

\end{abstract}


\section{Introduction}
\label{intro}

The rapid neutron-capture process, or \rpro,
produced heavy elements
observed in the earliest generations of stars
and about half of the heavy elements found in the solar system.
A common view from theory is that
compact binary mergers
are a viable site of \rpro\ nucleosynthesis
(e.g., \citealt{rosswog14,grossman14,thielemann17,wehmeyer19,
fernandez20,farouqi22}).
Rare kinds of supernovae may also be viable sites
to contribute at least some \rpro\ elements
(e.g., \citealt{winteler12,nishimura15,siegel19,fischer20,yong21nature}),
although normal core-collapse supernovae are not
(e.g., \citealt{fischer10,roberts10,arcones13}).
Recent observations generally do not contradict
the theoretical view.
The kilonova emission observed following the merger of 
two neutron stars detected in gravitational waves
appears to have been powered by the radioactive decay of 
a few $10^{-2}$~\msun\ of freshly produced \rpro\ elements
(e.g., \citealt{drout17,pian17,tanvir17}).
The concentration of the short-lived radioactive \rpro\ isotope
$^{244}$Pu in deep-sea sediments is 
consistent with a rare \rpro\ source with high yields,
such as neutron star mergers
\citep{hotokezaka15,wallner15,wallner21}.
A small fraction ($\approx$10\%) of the lowest-mass dwarf galaxies
around the Milky Way exhibit high levels of \rpro\ enhancement
among most of their stars
\citep{ji16nat,roederer16b,hansen17tuc3},
while all other galaxies in this mass range are extremely
deficient in \rpro\ elements (e.g., \citealt{frebel10ufd,ji19gru1tri2}).
Collectively these observations suggest that source(s)
of \rpro\ elements are rare but prolific events.
Nevertheless, multiple sites may be required
(e.g., \citealt{cote19,skuladottir20}).

The detailed element-by-element composition 
of \rpro-enriched ejecta
can be a powerful tool to distinguish among 
these candidates.
Kilonova spectra offer, at best, limited prospects for
measuring the detailed composition
of ejected \rpro\ material
\citep{zhu18,watson19,wu19}.
Uncertainties in calculating the atomic \citep{kasen13,fontes20}
and nuclear \citep{barnes21,zhu21}
data needed to interpret
kilonova spectra are large.
Old stars, on the other hand,
retain heavy-element signatures dominated by
individual \rpro\ events.
For example, 63 metals, including 42 \rpro\ elements,
have been detected in the 
metal-poor star \hdtwo\ \citep{roederer18c,roederer22a},
which presents the most complete chemical inventory known
for any object beyond the solar system.

These detailed chemical-abundance patterns
provide a key link between stellar
and nuclear astrophysics.
Closed neutron shells relevant to the \rpro\ occur in
nuclei with neutron numbers $N = 50$, 82, and 126.
Their neutron-capture cross sections are reduced relative to
neighboring nuclei, so they
produce peaks in the abundance distribution.
These peaks, at mass numbers $A \sim 80$, 130, and 195,
are sensitive probes of the physics of the \rpro\
(e.g., \citealt{kratz93,dillmann03,panov08,eichler15,shibagaki16,reiter20}).

Many \rpro-enhanced stars
exhibit a consistent relative abundance pattern
among the heavier \rpro\ elements,
here defined to be those with $56 \leq Z \leq 79$,
including the lanthanide and third-peak elements.
This pattern is a close match to the 
predicted solar system \rpro\ residual pattern.
This agreement implies that a robust \rpro\ 
operates across the history of the Galaxy,
a concept sometimes referred to as the ``universality'' of the \rpro.
Discussion of this phenomenon has appeared in
\citet{westin00} and 
many reviews over the years, including those by
\citet{sneden08}, \citet{frebel18}, and \citet{cowan21}.

The universal abundance pattern 
observed among the heavier \rpro\ elements
does not extend uninterrupted 
to the abundances of the lighter \rpro\ elements.
Many studies have shown that the
abundances of lighter \rpro\ elements,
here defined to be those with $34 \leq Z < 56$,
including the first- and second-peak elements,
are less correlated with the
abundances of heavier \rpro\ elements
(e.g., \citealt{wasserburg96,mcwilliam98,johnson02rpro,
travaglio04,francois07,qian08,hansen12}). 
Studies of the similarities of the abundance ratios among
strontium (Sr, $Z = 38$), 
yttrium (Y, $Z = 39$), 
zirconium (Zr, $Z = 40$), and other lighter \rpro\ elements
exist
(e.g., \citealt{aoki05,ivans06,hansen11pd,hansen14moru,wu15,aoki17,
cain18,roederer18c,spite18a}),
but generally they have not reported an analogous 
detection of universality among the abundances of the
lighter \rpro\ elements.

Theory predicts that
many nucleosynthesis processes,
in addition to the \rpro,
could potentially
contribute to the abundances of these 
lighter \rpro\ elements in the early Galaxy.
These processes include
the slow neutron-capture process (\spro)
in fast rotating massive stars 
(e.g., \citealt{pignatari08,pignatari10,cescutti13,choplin18,limongi18}),
the intermediate neutron-capture process (\ipro)
(e.g., \citealt{roederer16c,banerjee18,clarkson18}), and
proton-rich neutrino-driven wind components from core-collapse supernovae 
(e.g., \citealt{frohlich06,arcones11lepp,wanajo18b}).
The occurrence frequencies and the relative importance
of these different stellar processes are matter of debate. 

Here we contribute to this debate by
exploring the star-to-star dispersion among
the abundances of 12 lighter \rpro\ elements.
Our analysis begins with 
selenium (Se, $Z = 34$),
the lightest element with a substantial \rpro\ contribution \citep{roederer22a},
and concludes at 
tellurium (Te, $Z = 52$),
the heaviest element detectable 
with a mass less than 
barium (Ba, $Z = 56$) and the lanthanides.
We introduce our stellar sample and
present the abundances from the literature
in Section~\ref{data}.
We analyze these abundances collectively for the first time
in Section~\ref{results}.
We discuss the implications of our results
in Section~\ref{discussion}, and 
we summarize our conclusions
in Section~\ref{conclusions}.
We also present a minor update of abundances
derived from ultraviolet (UV) spectra,
motivated mainly by recent advances in the 
availability of atomic data,
in Appendices~\ref{appendix1}--\ref{appendix3}.

\section{Data}
\label{data}

Elements at the three \rpro\ peaks are readily
detectable in UV spectra 
($\lambda < 3100$~\AA)
collected using spectrographs
on the Hubble Space Telescope.
These elements include 
Se at the first peak
(e.g., \citealt{roederer12c,roederer12b,roederer14c,peterson20});
Te at the second peak
(e.g., \citealt{roederer12a,roederer12d,roederer16c,roederer22a});
and 
osmium (Os, $Z = 76$), iridium (Ir, $Z = 77$), and platinum (Pt, $Z = 78$)
at the third peak
(e.g., \citealt{cowan96,cowan05,sneden98,denhartog05,barbuy11}).
Se and Te are more difficult to detect than Os, Ir, or Pt,
because the Se~\textsc{i} and Te~\textsc{i} lines
are found at shorter UV wavelengths 
($2000 < \lambda < 2400$~\AA)
where longer exposure times are required to 
acquire sufficient signal-to-noise ratios.
In practice, 
these observations are limited to only very bright stars
with GALEX $NUV \lesssim 13.5$, or
Johnson $V \lesssim 9$ for stars with
effective temperature $>$~5500~K.

Our sample includes all metal-poor stars
with a Se or Te detection reported in the literature,
and whose lanthanide elements
exhibit abundance ratios dominated by the \rpro;
i.e., conform to the solar system \rpro\ residual pattern.
Only eight stars meet these criteria.
A summary of their names, 
metallicities ([Fe/H]),
abundance ratios of elements at
the three \rpro\ peaks
([Se/Fe], [Te/Fe], and [Pt/Fe]),
and regions between the peaks
([Zr/Fe] and [Eu/Fe])
are listed in Table~\ref{littab}.
A complete list of the adopted abundances,
including the adopted solar abundances,
is presented in
Appendix~\ref{appendix4}.
These eight stars span a range of 
metallicities and \rpro\ abundance ratios:\
$\approx$1.1~dex in [Fe/H],
$\approx$0.7~dex in [Zr/Fe], and 
$\approx$1.5~dex in [Eu/Fe].
The sample includes two highly \rpro-enhanced stars
([Eu/Fe] $> +0.7$, as defined by \citealt{holmbeck20}),
four moderately \rpro-enhanced stars
($+0.3 <$ [Eu/Fe] $\leq +0.7$), and
two stars with subsolar [Eu/Fe] ratios.
One star, \hdonefour,
satisfies the classification criteria 
proposed by \citet{frebel18}
for the ``weak'' or ``limited'' \rpro:\
[Eu/Fe] $< +0.3$, 
[Sr/Ba] $> +0.5$, and 
[Sr/Eu] $> 0.0$.
The complete abundance patterns for these stars
are available from
the references listed in Table~\ref{littab}.
We update the abundances to a common \loggf\ scale,
with references given in \citet{roederer22a}.

%\startlongtable
\begin{deluxetable*}{cccccccc}
\tablecaption{Stellar Sample, Sorted by Decreasing [Eu/Fe] Ratios
\label{littab}}
%\tablewidth{0pt}
\tabletypesize{\small}
\tablehead{
\colhead{Star} & 
\colhead{[Fe/H]} &
\colhead{[Se/Fe]} &
\colhead{[Zr/Fe]} &
\colhead{[Te/Fe]} &
\colhead{[Eu/Fe]} &
\colhead{[Pt/Fe]} &
\colhead{References}
}
\startdata
%\multicolumn{8}{c}{UV Sample} \\
%\hline
HD 222925   & $-1.46 \pm 0.10$ & $+0.74 \pm 0.22$ & $+0.62 \pm 0.08$ & $+0.91 \pm 0.14$ & $+1.32 \pm 0.08$ & $+1.29 \pm 0.10$ & 1 \\
\bd\        & $-2.10 \pm 0.20$ & \nodata          & $+0.35 \pm 0.14$ & $+0.34 \pm 0.30$ & $+0.90 \pm 0.04$ & $+1.01 \pm 0.07$ & 2, 3, 4, 5 \\
HD 108317   & $-2.37 \pm 0.14$ & $+0.47 \pm 0.42$ & $+0.24 \pm 0.20$ & $+0.41 \pm 0.30$ & $+0.48 \pm 0.18$ & $+0.51 \pm 0.19$ & 6, 7, 8 \\
HD 160617   & $-1.77 \pm 0.29$ & $+0.14 \pm 0.21$ & $+0.24 \pm 0.30$ & $+0.41 \pm 0.32$ & $+0.44 \pm 0.29$ & $+0.74 \pm 0.21$ & 9\tablenotemark{a}, 10 \\
HD 84937    & $-2.25 \pm 0.10$ & $+0.14 \pm 0.20$ & $+0.31 \pm 0.10$ & $+0.40 \pm 0.15$ & $+0.38 \pm 0.15$ & $< +0.53$        & 10, 11 \\
HD 19445    & $-2.15 \pm 0.10$ & $+0.37 \pm 0.23$ & $+0.38 \pm 0.10$ & $+0.62 \pm 0.15$ & $+0.37 \pm 0.15$ & $< +0.73$        & 10, 11 \\
HD 128279   & $-2.46 \pm 0.14$ & $-0.31 \pm 0.36$ & $-0.12 \pm 0.20$ & $+0.18 \pm 0.30$ & $-0.02 \pm 0.18$ & $< +0.08$        & 6, 7, 8 \\
HD 140283   & $-2.57 \pm 0.10$ & $+0.37 \pm 0.20$ & $-0.07 \pm 0.10$ & $< +0.59$        & $-0.22 \pm 0.10$ & $< -0.45$        & 10, 11 \\
\enddata
\tablereferences{%
1 = \citet{roederer22a}; 
2 = \citet{cowan02};
3 = \citet{sneden09};
4 = \citet{denhartog05};
5 = Appendix~\ref{appendix1};
6 = \citet{roederer12d};
7 = \citet{roederer14d};
8 = Appendix~\ref{appendix3};
9 = \citet{roederer12b};
10 = \citet{peterson20};
11 = Appendix~\ref{appendix2}.
}
\tablecomments{%
%The abundance ratio of elements X and Y relative to the
%Solar ratio is defined as
[X/Y] $\equiv \log_{10} (N_{\rm X}/N_{\rm Y}) - \log_{10} (N_{\rm X}/N_{\rm Y})_{\odot}$
}
\tablenotetext{a}{%
The Ir abundance is considered an upper limit, following discussion in Ref.\ 7.
}
%\tablenotetext{b}{%
%Adopting the Sn abundance from Ref.\ 9, 
%normalized to the scale in Ref.\ 8
%through the [Sn/Fe] ratio.
% }
%\tablenotetext{c}{%
%Sn abundance renormalized to the \loggf\ scale recommended by the
%National Institute of Standards and Technology (NIST)
%Atomic Spectra Database (ASD).
% }
\end{deluxetable*}

Appendix~\ref{appendix4}
also lists the stellar parameters
for the stars in our sample.
They span a range of evolutionary states, from the
main sequence to the red horizontal branch.
Previous studies \citep{aoki10,preston06,roederer14e}
have established that the \rpro\ abundance pattern is
recognizable in stars of all such evolutionary states
and is not impacted by the internal changes
that lead to stellar evolution.

The stars in our sample were originally selected 
for observations with the Space Telescope Imaging Spectrograph
(STIS; \citealt{kimble98,woodgate98})
for heterogeneous reasons.
\hdtwo\ and \bd\ were selected on the basis of their high [Eu/Fe] ratios
(europium, $Z = 63$).
\hdonezero\ and \hdonetwo\ were selected because they share
similar stellar parameters but exhibit a moderate contrast in their
[Eu/Fe] ratios.
\hdone, \hdeight, \hdonefour, and \hdonesix\ were selected
without regard to their heavy-element abundances.
These stars do not represent a comprehensive sample of
heavy-element abundance patterns observed in metal-poor stars.
As we show, however, they exhibit
a high degree of similarity in abundances of elements
that played no role in their original selection for observations.

\section{Results}
\label{results}

\subsection{Abundance Behaviors}
\label{behavior}

\begin{figure}
\begin{center}
\includegraphics[angle=0,width=3.35in]{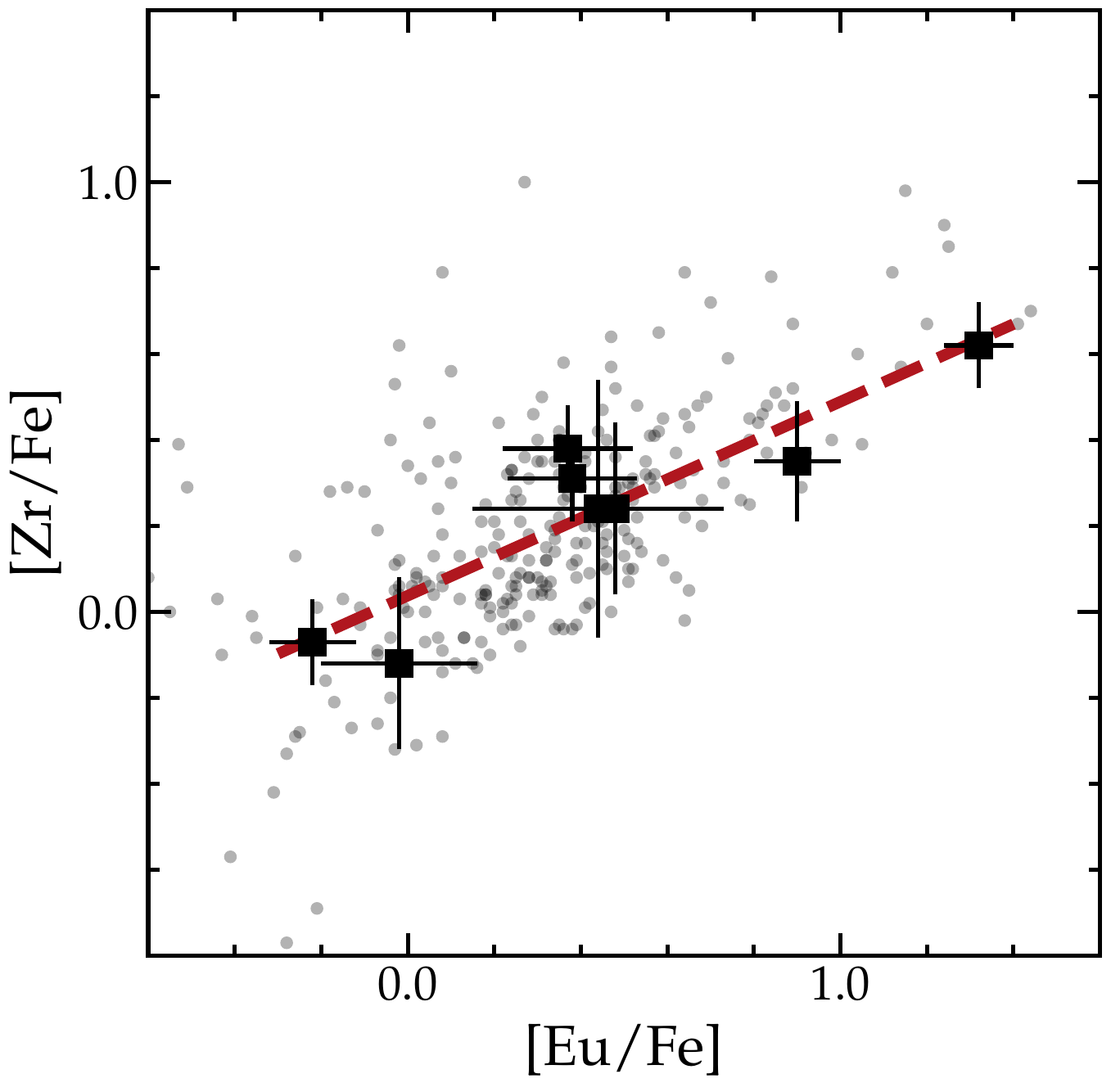}
\end{center}
\caption{
\label{zreuplot}
[Zr/Fe] versus [Eu/Fe] ratios in our sample (squares).
The red dashed line marks a linear fit to these eight stars.
A strong correlation with slope less than 1 is observed.
The small gray circles mark stars drawn from the literature
\citep{burris00,fulbright00,johnson02abund,honda04b,barklem05heres,
francois07,lai08,ishigaki13,roederer14c},
as found in JINAbase \citep{abohalima18}.
 }
\end{figure}

The [X/Fe] ratios
(for X = Se, Zr, Te, Eu, and Pt)
listed in Table~\ref{littab}
are correlated.
The Pearson correlation coefficients are $r > 0.65$ for 
each of these relationship pairs, 
indicating a high degree of correlation.
%
% Pearson correlation coefficients:
% X versus Zr:
%   Se: 0.71
%   Te: 0.88
%   Eu: 0.91
%   Pt: 0.92
% X versus Eu:
%   Se: 0.65
%   Zr: 0.91
%   Te: 0.75
%   Pt: 0.95
Figure~\ref{zreuplot} shows the
very high degree of correlation ($r = 0.91$)
between the [Zr/Fe] and [Eu/Fe] ratios.
The slope of the correlation is 
0.45, considerably less than 1.0.
In the analysis that follows, 
we account for this fact 
by scaling the
abundances of heavier \rpro\ elements to Eu 
and scaling the
abundances of lighter \rpro\ elements to Zr.
Zr and Eu are chosen because they
are detected in all eight stars,
are typically measured from multiple unsaturated lines
of the dominant ionization state,
and
their atomic transition probabilities are well known and do not dominate 
the uncertainty budget.

Figure~\ref{zreuplot} also demonstrates
that the stars in our sample provide a reasonable 
representation of the [Zr/Fe] and [Eu/Fe] ratios
found in metal-poor stars.
The small dots in Figure~\ref{zreuplot}
are drawn from literature samples
of metal-poor field stars that are not
enhanced in carbon or \spro\ elements.
Stars enhanced in \rpro\ elements are overrepresented 
in this comparison sample,
because it is easier to detect Zr and Eu when their
abundances are enhanced.
Nevertheless, despite the relatively limited availablity
of stars with low levels of Zr and Eu,
this comparison sample also
exhibits a similar correlation between
the [Zr/Fe] and [Eu/Fe] ratios.

\begin{figure*}
\begin{center}
\includegraphics[angle=0,width=3.35in]{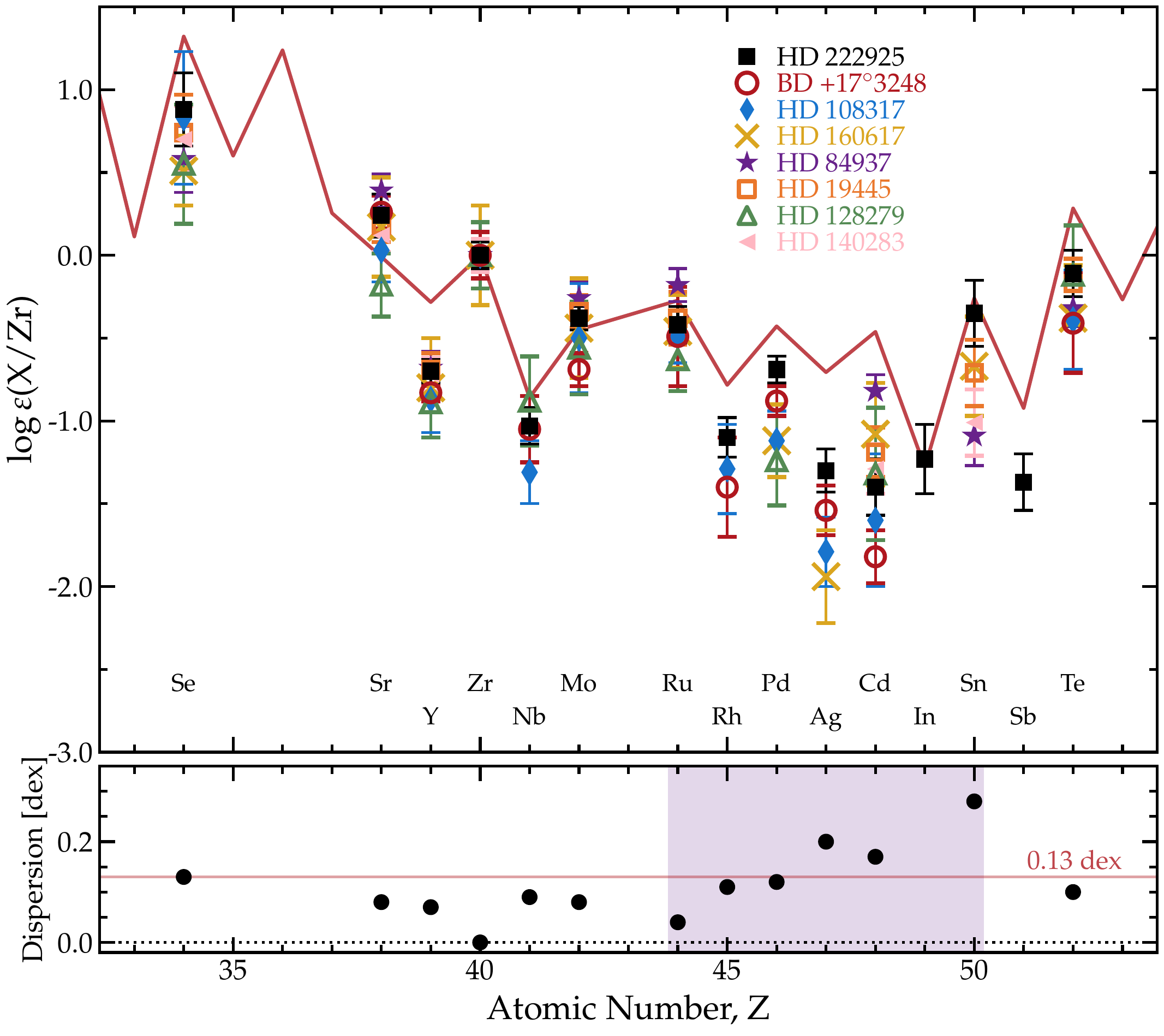}
\hspace*{0.1in}
\includegraphics[angle=0,width=3.35in]{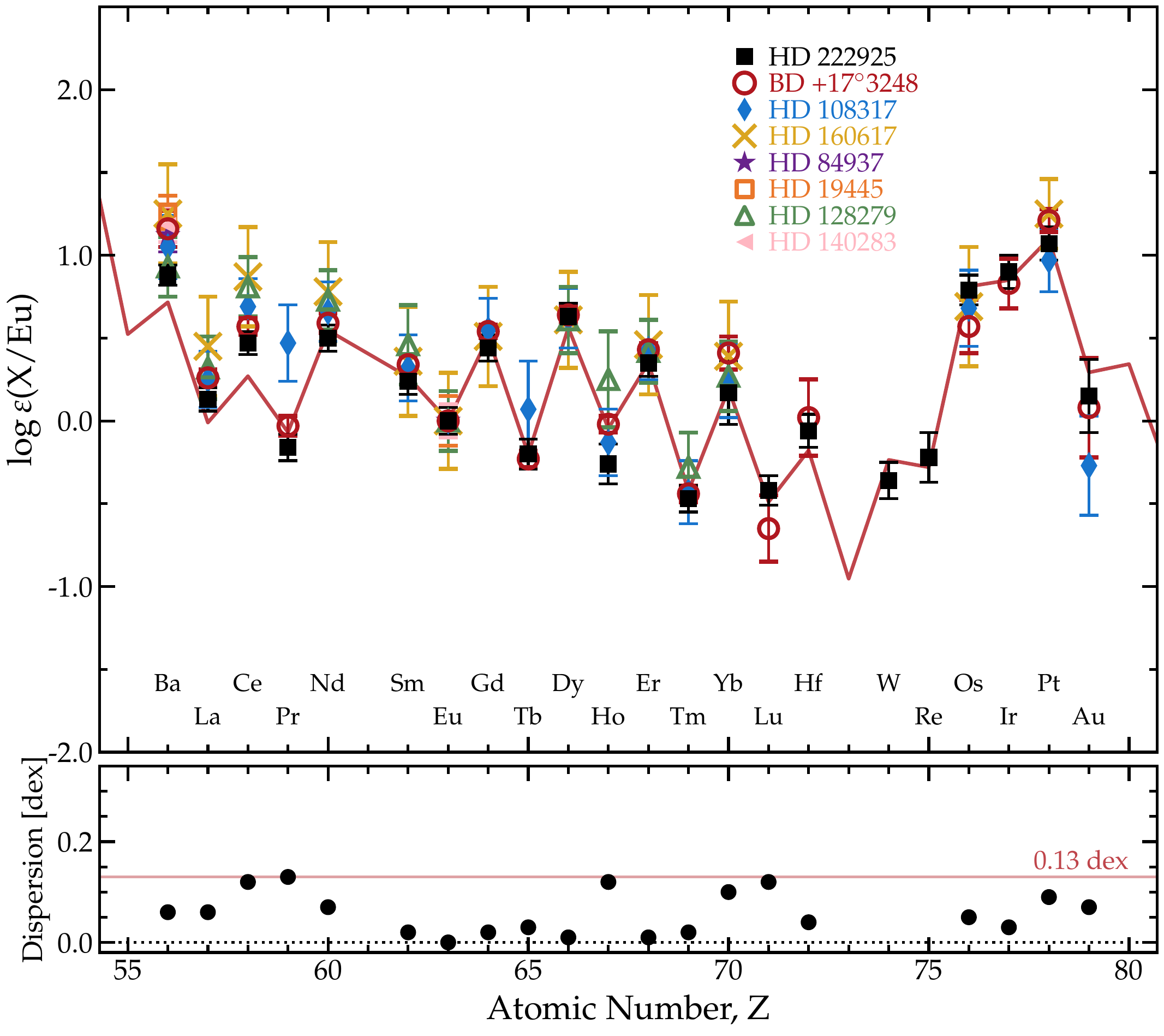}
\end{center}
\caption{
\label{abundplot}
Left:\ Abundances of lighter \rpro\ elements,
scaled to the Zr abundance in each star.
The legend is sorted by decreasing 
[Eu/Fe] ratios (Table~\ref{littab}).
Right:\ Abundances of heavier stable \rpro\ elements,
scaled to the Eu abundance in each star.
The red line in each panel marks the scaled solar system
\rpro\ residuals
\citep{prantzos20}.
The vertical axes span 4.5~dex in both panels.
The bottom panels illustrate the dispersions
of the \logeps{X/Zr} or \logeps{X/Eu} 
abundance ratios.
The dotted line in the bottom panels represents a dispersion of zero,
and the red lines represent dispersions of 0.13~dex.
The light purple box marks elements 
with conflicting abundance information,
or that may 
receive a substantial contribution from 
mechanisms beyond the scope of the present study
(Section~\ref{ruthenium}),
and they are not considered further here.
 }
\end{figure*}

Figure~\ref{abundplot} illustrates
the abundances of the stars in our sample.
The left panels show the lighter \rpro\ elements,
34 $\leq Z \leq$ 52,
scaled to the Zr
abundance in each star.
The right panels show the heavier \rpro\ elements,
56 $\leq Z \leq$ 79,
scaled to the Eu
abundance in each star.
The right panels of Figure~\ref{abundplot} 
illustrate the phenomenon 
commonly known as the universality of the \rpro.
The heavy elements in these stars
exhibit minimal star-to-star abundance dispersion when scaled
to account for the overall amount of \rpro\ material in each star.
We calculate the median absolute deviation (MAD) 
as a robust measure of the dispersion in each 
\logeps{X/Eu} ratio
(for X = Ba to Au).
These dispersions,
shown in the bottom-right panel of Figure~\ref{abundplot},
are typically small.
The abundances of most of the lanthanide elements 
are frequently derived from many unsaturated
and unblended lines in high-quality optical spectra.
The red line in the bottom panels of Figure~\ref{abundplot}
marks 0.13~dex (26\%), which is the upper boundary to the dispersion
among the lanthanide and third \rpro-peak element ratios.

Our choice to scale the abundances of lighter elements separately
reveals previously unrecognized abundance behavior.
A small star-to-star dispersion
is also observed among many elements
in the left panels of Figure~\ref{abundplot}.
The dispersions among the
\logeps{X/Zr} ratios
(for X = Se, Sr, Y, Nb, Mo, and Te)
are all small, $\leq$~0.13~dex (26\%), 
as shown in the bottom-left panel of Figure~\ref{abundplot}.
These values are also listed in Table~\ref{ratiotab},
along with the 
standard errors (std.\ err.)\ and
standard deviations (std.\ dev.).
These dispersions are comparable to the dispersions
among the heavier \rpro\ elements.
Critically, each individual
\logeps{X/Zr} ratio
(for X = Se, Sr, Y, Nb, Mo, and Te)
in each star
agrees to better than 2$\sigma$ with the
median calculated for the sample of eight stars.
Therefore, we propose that the abundance ratios among 
Se, Sr, Y, Zr, Nb, Mo, and Te
are also universal,
at the precision of available observational data,
provided that their abundances 
are scaled independently of the heavier elements.

%\startlongtable
\begin{deluxetable}{ccccc}
\tablecaption{Benchmark Abundance Ratios
\label{ratiotab}}
%\tablewidth{0pt}
\tabletypesize{\small}
\tablehead{
\colhead{X$_{1}$/X$_{2}$} &
\colhead{\logeps{X$_{1}$/X$_{2}$}} &
\colhead{std.\ err.} &
\colhead{std.\ dev.} &
\colhead{MAD} \\
\colhead{} &
\colhead{} &
\colhead{[dex]} &
\colhead{[dex]} &
\colhead{[dex]}
}
\startdata
Se/Zr &    0.70 & 0.10 & 0.13 & 0.13 \\
Sr/Zr &    0.20 & 0.06 & 0.16 & 0.08 \\
Y/Zr  & $-$0.73 & 0.06 & 0.08 & 0.07 \\
Nb/Zr & $-$1.06 & 0.10 & 0.16 & 0.08 \\
Mo/Zr & $-$0.42 & 0.06 & 0.13 & 0.06 \\
%Ru/Zr & $-$0.37 & 0.08 & 0.13 & 0.04 \\
Te/Zr & $-$0.23 & 0.08 & 0.12 & 0.10 \\
Se/Te &    0.93 & 0.13 & 0.16 & 0.05 \\
\enddata
%\tablereferences{%
%}
\tablecomments{%
%$\log\varepsilon$(X)~$\equiv \log_{10}(N_{\rm X}/N_{\rm H})+$12.0.
$\log\varepsilon$(X$_{1}$/X$_{2}$) 
$\equiv \log_{10}(N_{\rm X_{1}}/N_{\rm X_{2}})$
}
%\tablenotetext{a}{%
%}
\end{deluxetable}

Analysis of the abundance ratios found in a much larger sample
of metal-poor stars supports this conclusion.
Among the stars that are not classified
as carbon enhanced, \spro\ rich, \ipro\ rich, or $r+s$ in the
JINAbase abundance database \citep{abohalima18},
the MAD of the \logeps{Sr/Zr}, \logeps{Y/Zr}, and \logeps{Mo/Zr} ratios
are 0.14, 0.11, and 0.08~dex, respectively, based on samples of
294, 294, and 13~stars.
Dispersions among the Sr, Y, Zr, and Mo abundance ratios
for this larger sample are 
comparable to or smaller than our adopted 
0.13~dex dispersion criterion.
Less than 1.4\% of the stars in this sample
exhibit \logeps{X/Zr} (for X = Sr, Y, or Mo)
ratios that clearly deviate by more than 2$\sigma$ from the
median value of each ratio.
We regard the consistency in these ratios as superb,
considering the wide range of spectral quality and
inhomogeneous abundance analyses
reflected in the JINAbase sample.
This test suggests that the ratios among
Sr, Y, Zr, and Mo abundances
are similarly consistent in most metal-poor stars.

\subsection{Ruthenium through Tin}
\label{ruthenium}

Figure~\ref{abundplot} indicates that
the elements ruthenium (Ru, $Z = 44$), 
rhodium (Rh, $Z = 45$), and 
palladium (Pd, $Z = 46$)
also exhibit 
dispersions smaller than 0.13~dex in our sample.
The elements silver (Ag, $Z = 47$),
cadmium (Cd, $Z = 48$), and 
tin (Sn, $Z = 50$) exhibit dispersions
larger than 0.13~dex in our sample.
These elements are highlighted by the purple shaded box
in Figure~\ref{abundplot}.
We now discuss these elements' behavior in more detail.

Our Ru results, viewed in isolation,
favor the opposite conclusion than that
drawn by \citet{aoki17}.
That study, based on a sample of six stars,
concluded that the Ru/Zr ratios exhibit
significant star-to-star dispersion.
On one hand, the \logeps{Ru/Zr} ratios of the
stars in their sample agree to within 2$\sigma$,
which we would not consider to be a significant dispersion.
On the other hand,
\citeauthor{aoki17}\ also combined their sample with results from
\citet{sneden03a} (1~star) and 
\citet{hansen12,hansen14moru}
(12 giant stars with [Fe/H] $< -2.0$).
Collectively, the three samples exhibit a range of 
[Ru/Zr] ratios spanning $\approx$0.9~dex,
a fact that supported their finding of a significant dispersion 
among the Ru/Zr abundance ratios.
If we combine our sample with the \citeauthor{aoki17}\
expanded sample, we calculate MAD = 0.16~dex,
which we agree signals a significant dispersion in the Ru/Zr abundance ratios.
A similar calculation for the 
\logeps{Sr/Zr}, \logeps{Y/Zr}, and \logeps{Mo/Zr} ratios
in the \citeauthor{aoki17}\ expanded sample plus ours
yields MADs of 
0.09, 0.06, and 0.09~dex.
We thus confirm the results of \citeauthor{aoki17}\
that there are no significant differences among the abundances
of Sr, Y, Zr, and Mo.
We conclude that the Ru/Zr ratios may exhibit 
significant dispersion,
although our sample alone does not reveal such evidence.
A homogeneously analyzed larger sample of stars
may be necessary to 
definitively assess the abundance behavior of Ru.

The \logeps{X/Zr} ratios 
for X = Rh, Pd, Ag, Cd, and Sn
exhibit a moderate amount of dispersion in our sample,
$0.11 \leq$~MAD~$\leq 0.28$~dex.
\citet{vassh20} have proposed that 
these elements could receive 
a substantial contribution from
fission fragments
in some stars.
We explore the abundance behaviors of these elements separately
(Roederer et al., in preparation),
and so we do not discuss them further here.

\subsection{Tellurium}
\label{tellurium}

\begin{figure}
\begin{center}
\includegraphics[angle=0,width=3.35in]{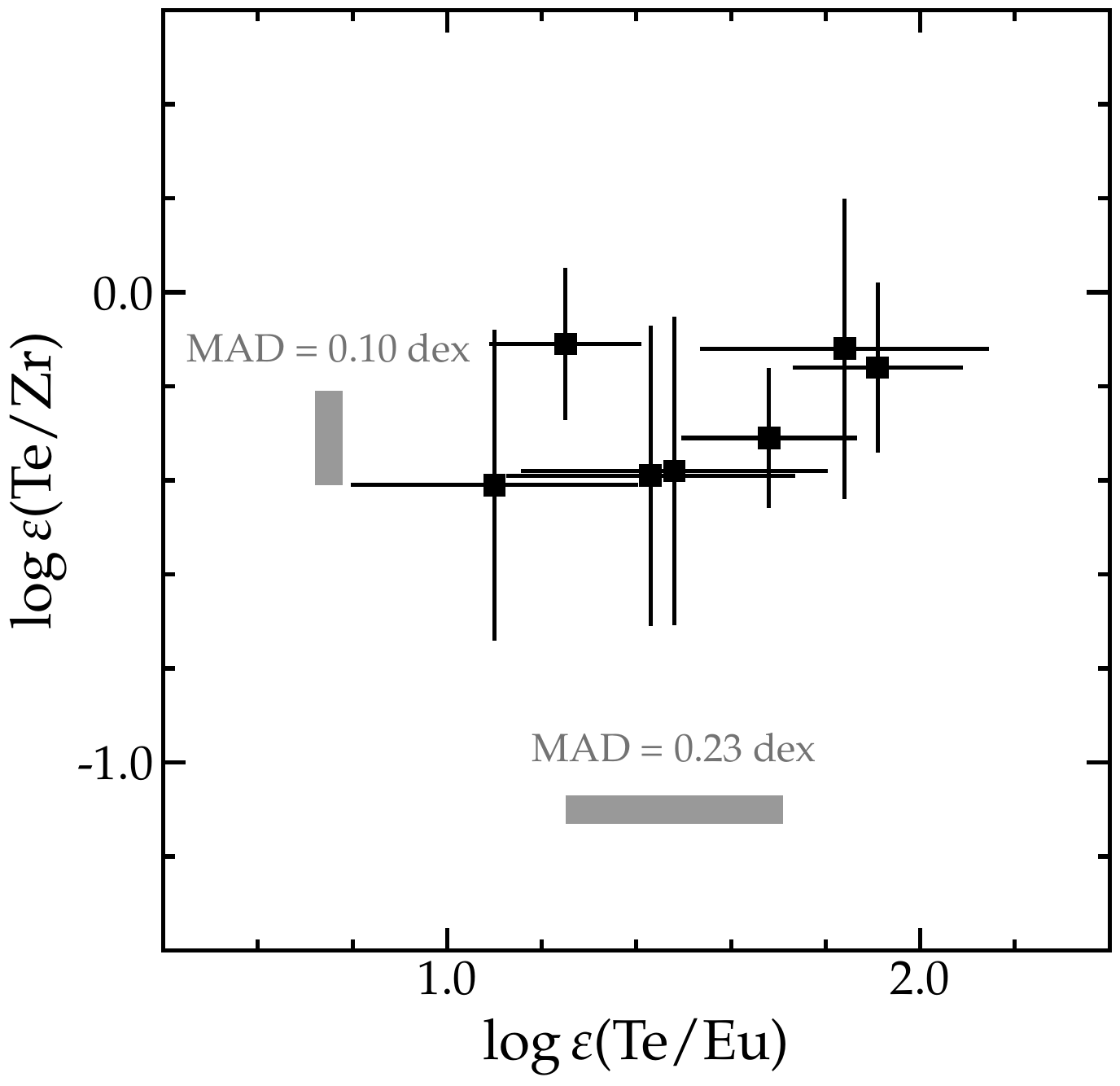}
\end{center}
\caption{
\label{teplot}
\logeps{Te/Zr} versus \logeps{Te/Eu} ratios
in our sample (black squares).
The shaded bars illustrate the 
median $\pm$ MAD for each ratio.
The MAD of the \logeps{Te/Zr} ratio is much smaller than that
of the \logeps{Te/Eu} ratio.
 }
\end{figure}

Te isotopes occupy the mass region
that could mark the transition
between the ``lighter'' and ``heavier'' \rpro\ elements,
so we discuss the Te abundances further in this section.
The Te abundances in our sample
correlate more strongly with the lighter elements 
than the heavier ones, 
as shown in Figure~\ref{teplot}.
The MAD and standard deviation
of the \logeps{Te/Zr} ratios are only 
0.10~dex and 0.12~dex,
while 
the MAD and standard deviation 
of the \logeps{Te/Eu} ratios are much larger, 
0.23~dex and 0.28~dex.
These differences in the dispersions 
cannot be due to differing observational uncertainties,
because
the median uncertainty in the 
Zr abundances, 0.10~dex,
is actually larger than
the median uncertainty in the
Eu abundances, 0.06~dex.
Te associates more naturally with
the lighter \rpro\ elements than the heavier ones
in this sample of stars.

\citet{roederer12a} reported that
Te abundances match the solar \rpro\ residual pattern
when scaled to Eu.
Our result supersedes that one.
The mean \logeps{Te/Eu} ratio in this sample is 
$1.52 \pm 0.09$ ($\sigma = 0.28$~dex),
whereas the solar system \rpro\ residuals
from \citet{goriely99}, \citet{sneden08}, \citet{bisterzo11,bisterzo14},
and \citet{prantzos20}
predict
\logeps{Te/Eu}$_{\odot,r} = 1.60$, 1.64, 1.62, and 1.60, respectively.
The Te abundances in this particular sample of stars
are low relative to the solar \rpro\ residual pattern
by only $\approx$0.1~dex, which is not significant.

We emphasize, however, that the Te abundances 
are more closely correlated with the
lighter \rpro\ elements,
such as Zr,
than the heavier \rpro\ elements, such as Eu.
We conclude that the transition from
the lighter to the heavier \rpro\ elements
likely occurs between Te and Ba,
around mass numbers 130 $\lesssim A \lesssim$ 135.

\subsection{Observational Benchmarks}
\label{benchmarks}

The weighted mean \logeps{X/Zr} ratios
for the elements that
do not exhibit significant star-to-star dispersion
(X = Se, Sr, Y, Nb, Mo, Te)
are listed in Table~\ref{ratiotab}.
Table~\ref{ratiotab} also lists the
weighted mean \logeps{Se/Te} ratio
as a measure of the relative heights of the
first and second \rpro\ peaks.

We compare a few of these key ratios with
predictions for the solar system \rpro\ residuals.
The mean \logeps{Se/Zr} ratio is 
$0.70 \pm 0.10$.
The solar system \rpro\ residuals
from \citet{goriely99}, \citet{sneden08}, \citet{bisterzo11,bisterzo14}, 
and \citet{prantzos20}
predict
\logeps{Se/Zr}$_{\odot,r} = 1.12$, 0.89, 1.22, and 1.33, respectively,
which are higher by
$\approx$0.2--0.6~dex than the ratios found
in our sample of metal-poor stars.
The mean \logeps{Te/Zr} ratio is 
$-0.23 \pm 0.08$.
The solar system \rpro\ residuals
from the same four references
predict
\logeps{Te/Zr}$_{\odot,r} = 0.12$, 0.29, 0.03, and 0.28, respectively,
which are higher by $\approx$0.3--0.5~dex
than the ratios found in our sample of metal-poor stars.
The mean \logeps{Se/Te} ratio is 
$0.93 \pm 0.13$.
This value is within the range of 
the solar system \rpro\ residual predictions
from the same four references, 
\logeps{Se/Te}$_{\odot,r} = 1.00$, 0.60, 1.19, and 1.05, respectively.
The values listed in Table~\ref{ratiotab} 
offer empirical \rpro\ abundance ratios
that are independent of the solar abundances.

Our proposed application of the concept of universality
(Section~\ref{behavior})
does not require a match to the solar system \rpro\ residual pattern,
as in the case of the heavier \rpro\ elements, because
the \rpro\ residual pattern for lighter elements
may be compromised.
Its derivation assumes
that the solar system abundance pattern
contains material from only
the \rpro, \spro,
and proton-capture process (\textit{p}-process).
This key assumption is likely invalid
(e.g., \citealt{travaglio04,bisterzo17,cote18ipro}).
Thus, the fair but imperfect agreement between the
\rpro\ residual pattern and the 
stellar Se, Sr, Y, Zr, Nb, Mo, and Te abundances
does not contradict
the operation of a robust \rpro\ 
across cosmic time.
If multiple \rpro\ sources operate, 
their occurrence frequencies may have been different
in the early Galaxy than they are in the Galactic disk
(e.g., \citealt{cote19}).
This scenario could offer one explanation 
for the differences between the 
solar system \rpro\ residual pattern and the
benchmark abundance pattern presented here.

\subsection{Comparison with Other Classes of \rpro-Enhanced Stars}
\label{other}

In this section we search for differences between
the lighter \rpro-element abundance pattern found in our sample
and that found in two other classes of stars.
One class is the \rpro-enhanced stars with 
the so-called actinide boost,
characterized by high Th/Eu ratios,
here defined to be \logeps{Th/Eu} $> -0.40$,
relative to most other \rpro-enhanced stars.
This class includes 11~stars:\
\object[BPS CS 22953-003]{CS~22953--003} \citep{roederer14c},
\object[BPS CS 30306-132]{CS~30306--132} \citep{honda04b},
\object[BPS CS 30315-029]{CS~30315--029} \citep{siqueiramello14},
\object[BPS CS 31078-018]{CS~31078--018} \citep{lai08},
\object[BPS CS 31082-001]{CS~31082--001} \citep{siqueiramello13},
\object[HD 6268]{HD~6268} \citep{roederer14c}, 
\object[HE 1219-0312]{HE~1219$-$0312} \citep{hayek09},
\object[HE 2252-4225]{HE~2252$-$4225} \citep{mashonkina14he},
\object[HE 2327-5642]{HE~2327$-$5642} \citep{mashonkina10},
\object[2MASS J09544277+5246414]{J0954$+$5246} \citep{holmbeck18}, and
\object[SMSS J200322.54-114203.3]{SMSS~J2003$-$1142} \citep{yong21nature}.
Abundances for Sr, Y, and Zr are reported for each star,
but abundances for Nb and Mo are only available for 
one and six stars in this set, respectively.
None of the \logeps{X/Zr} ratios (for X = Sr, Y, Nb, or Mo)
in these stars
differ by more than 2$\sigma$ from the mean benchmark ratios.

The other class of stars is the
so-called weak (or limited) \rpro\ stars.
It is exemplified by two stars examined by \citet{honda06,honda07},
\object[HD 88609]{HD~88609} and 
\object[HD 122563]{HD~122563},
which exhibit
relatively low enhancements of the elements beyond the second
\rpro\ peak.
Abundances of Sr, Y, Zr, Nb, and Mo
are available for these two stars.
None of the \logeps{X/Zr} ratios in these stars
differs by more than 2$\sigma$ from the mean benchmark ratios,
with the exception of the \logeps{Nb/Zr} ratio in 
\object[HD 88609]{HD~88609}, which is 
low by only $\approx$2.3$\sigma$.

We conclude from these comparisons that
neither class of stars
exhibits significantly different abundance behavior
among the elements from Sr through Mo.
It would be worthwhile to perform similar comparisons
for the Se and Te abundances,
which, unfortunately, 
are not available at present
for any of the stars in these classes.

\subsection{The Impact of Potential Contamination}
\label{contamination}

We assume that the abundance patterns in these stars
reflect only \rpro\ nucleosynthesis.
Other processes 
could contribute to the heavy elements in these stars,
in principle,
as discussed in Section~\ref{intro}.
We construct a toy model to test our assumption
using the \logeps{Se/Zr} and \logeps{Te/Zr} ratios.
We explore the consequences of \spro\ and \ipro\ contamination
from short-lived massive stars
that could have enriched the 
gas from which the stars in our sample were formed.
For now, we do not distinguish between a weak (or limited) \rpro\
and a main \rpro,
and we defer this discussion to Section~\ref{discussion}.

We adopt a set of \spro\ abundance ratios representative of the production 
in fast-rotating massive stars 
(e.g., \citealt{pignatari08,frischknecht16,choplin18,limongi18}).
We use a single-zone trajectory from a 25~\msun\ star 
with an initial metallicity of $Z = 10^{-4}$ ([Fe/H] = $-2.14$),
calculated using the Geneva stellar evolution code GENEC \citep{hirschi04}.
This model is representative of stars with \spro\ production during both 
core He-burning and shell C-burning \citep{nishimura17}. 
We perform these nucleosynthesis calculations using the
post-processing network code PPN \citep{pignatari12}.
We adopt the parametric approach used by \citet{pignatari13}
to account for the possible range of \spro\ production,
by varying the amount of primary $^{22}$Ne 
destroyed to make neutrons via the $^{22}$Ne($\alpha$,n)$^{25}$Mg reaction.
Nine sets of \spro\ ratios,
ranging from 
$-0.74 \leq$ \logeps{Se/Zr} $\leq$ 1.62
and
$-2.43 \leq$ \logeps{Te/Zr} $\leq -0.96$,
are considered.

One-dimensional massive star models incorporating \ipro\ calculations exist,
but they await verification by comprehensive 
hydrodynamics simulations that are able to resolve
and capture the main properties of these types of events 
(e.g., \citealt{herwig14,woodward15}).
Accordingly, we adopt a simplified set of \ipro\ conditions 
that qualitatively capture the main properties of \ipro\ abundance patterns 
with different amounts of neutrons.
We use the nucleosynthesis framework 
adopted in \citet{bertolli13} and \citet{roederer16c}
for a metallicity of [Fe/H] = $-2.2$.
Seven \ipro\ neutron exposures
are considered, producing abundance ratios that range from
$-1.76 \leq$ \logeps{Se/Zr} $\leq$ 1.67
and
$-1.35 \leq$ \logeps{Te/Zr} $\leq$ 1.20.

Finally, we adopt \rpro\ abundance ratios
from the observationally derived values presented in
Table~\ref{ratiotab}.

We calculate seven realizations of the 
\logeps{Se/Zr} and \logeps{Te/Zr} ratios
in each simulation, matching the number of stars in our sample
where each ratio is derived.
We adjust the mean relative mix of material from the different 
neutron-capture processes.
Each component of the input mixture is multiplied by 
a random number between 0 and 1 in each simulation
to account for variability in the mixtures,
although the overall mean relative mix is maintained
in each simulation.
We also inject observational uncertainties into each of these ratios.
The mean (median) 
observational uncertainties in the Se and Te 
abundances are
0.26~dex (0.22~dex) and
0.24~dex (0.30~dex), respectively, 
so we apply a generously small 0.2~dex observational uncertainty
to our resampled abundances.
We then calculate the standard deviation
of the seven resamples of each ratio.
We repeat this process $10^{4}$ times.

We conduct simulations with
95/5, 80/20, and 50/50 \%
mean mixes between the \rpro\ ratios and each of the
\spro\ and \ipro\ ratios.
We also conduct a simulation with equal mixes of all three sets of ratios.
Fewer than 5\% of these simulations
in any given mixture exhibit equal or less dispersion
than our sample, 
and the percentage is $<$~2\% for the
majority of these simulations.
This unsurprising result indicates that the inclusion
of material from other processes 
can only increase the dispersion in the observed
abundance ratios.

Alternatively, contamination could bias the means
of these ratios.
We consider this possibility unlikely.
The dominance of the \rpro\ that contributed 
heavy elements to \hdtwo,
which has the most \rpro\ material
among stars in our sample 
([Eu/Fe] = $+1.32$, \logeps{Eu} = 0.38), 
masks any contributions from other sources to \hdtwo.
However,
this star exhibits statistically indistinguishable abundance ratios
among Se, Sr through Mo, and Te 
when compared to stars with much lower levels of
\rpro\ enhancement 
($-0.22 \leq$ [Eu/Fe] $\leq +0.90$)
or enrichment
($-2.27 \leq$ \logeps{Eu} $\leq -0.68$).
Any bias from contamination
is smaller than 0.13~dex (26\%),
on average.

\section{Discussion}
\label{discussion}

For the eight stars examined in this work,
the abundances of the heavy \rpro\ elements 
exhibit ratios constant to within a dispersion of 0.13~dex,
roughly equivalent to observational precision.
This uniformity persists despite the [Eu/Fe] ratios
changing by a factor of more than 30,
a fact established by previous work
(e.g., \citealt{spite18a}).
We demonstrate,
using the same stars,
that the \rpro\ abundance ratios among the
elements Se, Sr, Y, Zr, Nb, Mo, and Te
exhibit star-to-star dispersion comparable to
that exhibited by the
lanthanides and, when detected, the third-peak elements.
We propose that, in this case,
the universality of these lighter \rpro\ elements
at and between the first and second \rpro\ peaks
mirrors the universality of the \rpro\ for the
lanthanides through the third peak,
provided that the abundances of the lighter \rpro\ elements
are scaled independently of the heavier ones.

One possible interpretation of this behavior is
to associate the weak (or limited) \rpro\ with the lighter \rpro\ elements
and the main \rpro\ with the heavier \rpro\ elements.
Several theoretical studies have used \rpro\ nucleosynthesis models
to explore the transition region between these two sets of elements
(e.g., \citealt{kratz07,montes07,farouqi10,wanajo13,lorusso15,wu16}).
\citeauthor{kratz07}, for example, 
found that a weak \rpro\ with neutron densities up to
$\log n_{n} \lesssim 10^{23}$ 
could produce the lighter elements, 
including a peak around $A \sim 130$.
The \rpro\ reaction network calculations of \citeauthor{lorusso15},
which incorporated newly measured $\beta$-decay halflives of
110 neutron-rich nuclei leading to the second-peak region,
indicated that the robust nature of Te production
differed from that of the lanthanides.
This boundary between the weak and main {\rpro}es likely
behaves as a transition, not a cutoff, 
and some amount of Te and second-peak elements 
will also be produced by 
higher neutron densities characteristic of the main \rpro.

Despite the multitude of processes
that could theoretically produce elements
with $A \lesssim 130$,
no star in our sample exhibits
abundance ratios 
among Se, Sr through Mo, and Te
that are significantly distinct from the others.
If multiple \rpro\ sites or other processes contribute to the
inventory of lighter \rpro\ elements incorporated into
metal-poor stars dating from the early Universe,
they must occur relatively infrequently,
eject relatively low yields,
produce abundance ratios that are
not distinct from those listed in Table~\ref{ratiotab},
or some combination of these scenarios.
We challenge theorists to 
identify which \rpro\ sites
are capable of producing 
both the lighter and the heavier \rpro\ elements
and producing robust abundance
distributions that match the elemental abundance patterns.

Our sample is small and limited by 
access to high-resolution UV spectra.
A few stars in much larger optical-only samples 
may exhibit deviant 
\logeps{Sr/Zr} or \logeps{Y/Zr} ratios
(e.g., \citealt{lombardo22}).
It will be of interest to evaluate whether such abundance
ratios reflect true deviations from the \rpro\ or 
contributions from another, presumably more rare, process.
We challenge observers to 
expand the sample of stars with high-quality
abundance derivations for lighter \rpro\ elements,
identify any stars whose lighter \rpro-element abundance ratios
deviate significantly from the pattern we have characterized,
and establish the range of 
\logeps{X/Zr} ratios that may occur in metal-poor stars.

\section{Conclusions}
\label{conclusions}

We examine the abundances of elements at and between the
first and second \rpro\ peaks
for eight stars with Se or Te detections.
Our data are drawn from the literature,
where each study typically focused on one or a few stars
at a time.
Here, we aggregate these data to
characterize the abundance patterns and behaviors
of elements from Se through Te for the first time.

The abundances of the lighter \rpro\ elements
vary relative to
the yields of the heavier \rpro\ elements,
but they are not fully decoupled 
(Figure~\ref{zreuplot}).
For this sample of stars, 
many of the lighter \rpro\ elements, including
Se, Sr, Y, Zr, Nb, Mo, and Te
exhibit abundance ratios
with dispersions
$\leq 0.13$~dex (26\%; Figure~\ref{abundplot}).
This level of consistency 
matches that of heavier \rpro\ elements.
We propose that we have potentially identified a universality
analogous to that observed among the lanthanides and
third \rpro-peak elements,
at the precision of available observational data,
provided the overall abundances of lighter \rpro\ elements
are scaled independently of the heavier ones.
The abundance behavior of the elements from
Ru through Sn requires further study.
We calculate benchmark ratios among the abundances
of Se, Sr through Mo, and Te
(Table~\ref{ratiotab}).
These values reflect yields of events that enriched
the gas from which metal-poor stars formed.
They can be used to constrain models of
candidate sites of \rpro\ nucleosynthesis in the early Universe.
We conclude that
at least one relatively common \rpro\ source in the early Universe
produced a consistent abundance pattern
among some light elements spanning
the first and second \rpro\ peaks.

\acknowledgments

We thank J.\ Lawler and C.\ Sneden for 
thoughtful discussions
and the referee for helpful suggestions to improve this manuscript.
We %I.U.R., T.C.B., A.F., and V.M.P.\ 
acknowledge generous support 
provided by NASA through grants GO-14765, GO-15657, and GO-15951
from the Space Telescope Science Institute, 
which is operated by the Association of Universities 
for Research in Astronomy, Incorporated, 
under NASA contract NAS5-26555.
We acknowledge support 
awarded by the U.S.\ National Science Foundation (NSF):\
grants 
PHY~14-30152 (Physics Frontier Center/JINA-CEE),
OISE~1927130 
(International Research Network for Nuclear Astrophysics/IReNA),
AST~1716251 (A.F.), 
AST~1814512 (E.A.D.H.), and
AST~1815403/1815767 (I.U.R.).
I.U.R.\ acknowledges support from the NASA
Astrophysics Data Analysis Program, grant 80NSSC21K0627.
M.P.\ acknowledges significant support to NuGrid from STFC 
(through the University of Hull's Consolidated Grant ST/R000840/1),
and access to {\sc viper}, 
the University of Hull High Performance Computing Facility. 
M.P.\ also acknowledges support from the ``Lend\"{u}let-2014'' Programme 
of the Hungarian Academy of Sciences (Hungary), 
and from the ERC Consolidator Grant (Hungary) 
funding scheme (project RADIOSTAR, G.A.\ n.\ 724560). 
M.P.\ thanks the ChETEC COST Action (CA16117), 
supported by COST (European Cooperation in Science and Technology), 
and the ChETEC-INFRA project funded from the European Union's 
Horizon 2020 research and innovation programme 
(grant agreement No 101008324). 
The work of V.M.P.\ is supported by NOIRLab, 
which is managed by AURA under a cooperative agreement with 
the NSF.~
M.R.M.\ was supported by the 
Laboratory Directed Research and Development program 
of Los Alamos National Laboratory 
under project number 20200384ER.~
Los Alamos National Laboratory is operated by Triad National Security, LLC, 
for the National Nuclear Security Administration 
of U.S.\ Department of Energy (Contract No.\ 89233218CNA000001).
R.S.\ acknowledges support from grant
DE-FG02-95-ER40934
awarded by the U.S.\ Department of Energy.
This research has made use of NASA's
Astrophysics Data System Bibliographic Services;
the arXiv preprint server operated by Cornell University;
the SIMBAD and VizieR
databases hosted by the
Strasbourg Astronomical Data Center;
the ASD hosted by NIST;
the Mikulski Archive for Space Telescopes
at the Space Telescope Science Institute, which is 
operated by the Association of Universities for 
Research in Astronomy, Inc.\ under NASA contract NAS~5-26555;
and
the Keck Observatory Archive,
which is operated by the W.~M.\ Keck Observatory and 
the NASA Exoplanet Science Institute,
under contract with NASA. %; and
%Image Reduction and Analysis Facility (IRAF) software packages
%distributed by the National Optical Astronomy Observatories,
%which are operated by AURA,
%under cooperative agreement with the NSF.~

\facility{HST (STIS), Keck (HIRES)}

\software{%
%IRAF \citep{tody93},
LINEMAKE \citep{placco21linemake},
matplotlib \citep{hunter07},
MOOG \citep{sneden73,sobeck11},
numpy \citep{vanderwalt11},
scipy \citep{jones01}}

\appendix
\restartappendixnumbering

Recent advances in laboratory measurements and
theoretical calculations of atomic data
justify revision of abundances of heavy elements
that were published prior to the availability of these data.
In the following appendices, we present
revised abundances and, in a few cases,
newly derived abundances 
for a few metal-poor stars
with high-quality UV spectra.

\section{Updated Heavy-Element Abundances in BD~$+$17$^{\circ}$3248}
\label{appendix1}

\bd\ has been the subject of many detailed abundance studies
over the last two decades.
As a result of these many piecemeal reanalyses,
the heavy-element abundances for \bd\
are scattered throughout the literature.
Different versions 
of the local thermodynamic equilibrium % (LTE) 
analysis code MOOG \citep{sneden73} have been used
to derive these abundances.
Furthermore,
new high-quality laboratory atomic data have 
become available since the original publications.
Here, we present a homogeneous, revised set of 
abundances of 16 \rpro\ elements
in \bd\ to make use of these advances.

We use the STIS 
E230M spectrum of \bd\
($R$ = 30,000, 2280 $\leq \lambda \leq$ 3120~\AA,
GO-8342, datasets O5F607010-020, PI Cowan) 
to rederive abundances for all elements except Ag, 
for which we use the Keck High Resolution Echelle Spectrometer 
(HIRES; \citealt{vogt94}) 
spectrum obtained through the Keck Observatory Archives
($R$ = 45,000, 3120 $\leq \lambda \leq$ 4640~\AA;
U25H, PI Fuller).
All abundances listed in Table~\ref{linetab} 
are derived with the most recent version of MOOG,
which incorporates the treatment of Rayleigh scattering
as described in \citet{sobeck11}.
We use an ATLAS9 \citep{castelli04} model atmosphere interpolated
to the parameters derived by
\citet{cowan02},
effective temperature (\teff) = 5200 $\pm$~150~K,
log of the surface gravity (\logg) = 1.80 $\pm$~0.3,
metallicity ([Fe/H]) = $-2.10$ $\pm$~0.2, and
microturbulence velocity parameter (\vt) = 1.9 $\pm$~0.2~\kmsec.
We derive abundances by spectrum synthesis matching
using the MOOG ``synth'' driver,
with line lists generated using the LINEMAKE code \citep{placco21linemake}.
Table~\ref{abundtab} lists the recommended abundances.
Uncertainties are computed following the method described in
\citet{roederer18c}.
All other \rpro\ elements not listed in Table~\ref{abundtab}
should be adopted from the literature, in order
of decreasing priority in the case of duplicate values:\
\citet{sneden09}, 
\citet{denhartog05}, 
\citet{roederer10b}, or
\citet{cowan02}.

%\startlongtable
\begin{deluxetable}{cccccccc}
\tablecaption{Updated Abundances from UV Spectrum of BD $+$17$^{\circ}$3248
\label{linetab}}
%\tablewidth{0pt}
\tabletypesize{\small}
\tablehead{
\colhead{Species} &
\colhead{$\lambda$} &
\colhead{$E_{\rm low}$} &
\colhead{\loggf} &
\colhead{Ref.} &
\colhead{$\log\epsilon$} &
\colhead{Fit Unc.} &
\colhead{Supersedes} \\
\colhead{} &
\colhead{(\AA)} &
\colhead{(eV)} &
\colhead{} &
\colhead{} &
\colhead{} &
\colhead{[dex]} &
\colhead{} 
}
\startdata
Ge~\textsc{i}  &    2651.17 & 0.17 & $-$0.07 &  1 &    0.62 & 0.20 & \nodata \\
Ge~\textsc{i}  &    2691.34 & 0.07 & $-$0.81 &  1 &    0.42 & 0.20 & \nodata \\
Ge~\textsc{i}  &    3039.07 & 0.88 & $-$0.07 &  1 &    0.47 & 0.15 & 15 \\
As~\textsc{i}  &    2288.11 & 1.35 & $-$0.06 &  2 &$<-$0.10 &\nodata&\nodata \\
Nb~\textsc{ii} &    2910.59 & 0.38 & $-$0.16 &  3 & $-$0.21 & 0.15 & \nodata \\
Nb~\textsc{ii} &    2911.74 & 0.33 & $-$0.28 &  3 & $-$0.18 & 0.20 & \nodata \\
Mo~\textsc{ii} &    2660.58 & 1.49 & $-$0.14 &  4 &    0.18 & 0.15 & \nodata \\
Mo~\textsc{ii} &    2871.51 & 1.54 & $+$0.06 &  4 &    0.00 & 0.15 & \nodata \\
Mo~\textsc{ii} &    2911.92 & 1.60 & $-$0.10 &  4 &    0.30 & 0.15 & \nodata \\
Mo~\textsc{ii} &    2930.50 & 1.49 & $-$0.23 &  4 &    0.09 & 0.15 & \nodata \\
Ag~\textsc{i}  &    3280.68 & 0.00 & $-$0.02 &  5 & $-$0.86 & 0.20 & 8 \\
Ag~\textsc{i}  &    3382.89 & 0.00 & $-$0.33 &  5 & $-$0.62 & 0.15 & 8 \\
Cd~\textsc{i}  &    2288.02 & 0.00 & $+$0.15 &  6 & $-$0.99 & 0.15 & 8 \\
In~\textsc{ii} &    2306.06 & 0.00 & $-$2.30 &  2 &$<-$0.10 &\nodata&\nodata \\
Te~\textsc{i}  &    2385.79 & 0.59 & $-$0.81 &  7 &    0.42 & 0.30 & 7 \\
Lu~\textsc{ii} &    2615.41 & 0.00 & $+$0.11 &  8 &$<-$1.47 &\nodata&8 \\
Lu~\textsc{ii} &    2911.39 & 1.76 & $+$0.45 &  9 & $-$1.33 & 0.20 & \nodata \\
Hf~\textsc{ii} &    2322.48 & 0.00 & $-$1.14 & 10 & $-$1.10 & 0.20 & \nodata \\
Hf~\textsc{ii} &    2641.41 & 1.04 & $+$0.57 & 11 & $-$0.67 & 0.15 & \nodata \\
Hf~\textsc{ii} &    2861.70 & 0.45 & $-$0.32 & 10 & $-$1.14 & 0.20 & \nodata \\
Hf~\textsc{ii} &    3012.90 & 0.00 & $-$0.61 & 11 & $-$0.82 & 0.20 & \nodata \\
Os~\textsc{i}  &    3058.65 & 0.00 & $-$0.41 & 12 &    0.06 & 0.20 & 15 \\
Os~\textsc{ii} &    2282.28 & 0.00 & $-$0.14 & 13 & $-$0.39 & 0.15 & 8 \\
Ir~\textsc{i}  &    2639.71 & 0.00 & $-$0.31 & 14 & $-$0.11 & 0.15 & \nodata \\
Ir~\textsc{i}  &    2924.79 & 0.00 & $-$0.66 &  2 &    0.22 & 0.25 & \nodata \\
Au~\textsc{i}  &    2675.94 & 0.00 & $-$0.60 &  2 & $-$0.60 & 0.30 & 16 \\
\enddata
\tablereferences{%
 1 = \citet{li99};
 2 = \citet{roederer22a};
 3 = \citet{nilsson08};
 4 = \citet{sikstrom01};
 5 = \citet{hansen12};
 6 = \citet{morton00};
 7 = \citet{roederer12a};
 8 = \citet{roederer10b};
 9 = \citet{denhartog20};
10 = \citet{denhartog21hf};
11 = \citet{lawler07};
12 = \citet{quinet06};
13 = \citet{ivarsson04};
14 = NIST \citep{kramida21};
15 = \citet{cowan05};
16 = \citet{cowan02}.
}
%\tablecomments{%
%}
%\tablenotetext{a}{%
%}
\end{deluxetable}

%\startlongtable
\begin{deluxetable}{ccccccc}
\tablecaption{Updated Mean Abundances in BD~$+$17$^{\circ}$3248
\label{abundtab}}
%\tablewidth{0pt}
\tabletypesize{\small}
\tablehead{
\colhead{Element} &
\colhead{$\log\epsilon_{\odot}$\tablenotemark{a}} &
\colhead{$\log\epsilon$} &
\colhead{[X/Fe]\tablenotemark{b}} &
\colhead{Unc.} &
\colhead{N$_{\rm lines}$} &
\colhead{Notes} \\
\colhead{} &
\colhead{} &
\colhead{} &
\colhead{} &
\colhead{[dex]} &
\colhead{} &
\colhead{}
}
\startdata
Ge &    3.65 &    0.50 & $-$1.05 & 0.15 &  3 & \nodata \\
As &    2.30 &$<-$0.10 &$<-$0.30 &\nodata& 1 & \nodata \\
Sr &    2.87 &    1.09 & $+$0.32 & 0.10 &  2 & \tablenotemark{c} \\
Y  &    2.21 &    0.00 & $-$0.11 & 0.05 & 11 & \tablenotemark{d} \\
Zr &    2.58 &    0.83 & $+$0.35 & 0.14 & 19 & \tablenotemark{e} \\
Nb &    1.46 & $-$0.22 & $+$0.42 & 0.20 &  3 & \tablenotemark{f} \\
Mo &    1.88 &    0.14 & $+$0.36 & 0.10 &  5 & \tablenotemark{f} \\
Ag &    0.94 & $-$0.71 & $+$0.45 & 0.15 &  2 & \tablenotemark{g} \\
Cd &    1.71 & $-$0.99 & $-$0.60 & 0.16 &  1 & \nodata \\
In &    0.80 &$<-$0.10 &$<+$1.20 &\nodata& 1 & \nodata \\
Te &    2.18 &    0.42 & $+$0.34 & 0.30 &  1 & \nodata \\
Lu &    0.10 & $-$1.33 & $+$0.67 & 0.20 &  1 & \tablenotemark{h} \\
Hf &    0.85 & $-$0.66 & $+$0.59 & 0.23 & 10 & \tablenotemark{i} \\
Os &    1.40 & $-$0.11 & $+$0.59 & 0.16 &  3 & \tablenotemark{j} \\
Ir &    1.38 &    0.15 & $+$0.87 & 0.15 &  5 & \tablenotemark{k} \\
Au &    0.92 & $-$0.60 & $+$0.58 & 0.30 &  1 & \nodata \\
\enddata
%\tablecomments{%
%All abundances include ionization corrections.
%}
\tablenotetext{a}{%
\citet{asplund09}
}
\tablenotetext{b}{%
Referenced to [Fe/H] $= -$2.10
}
\tablenotetext{c}{%
Abundance from \citet{cowan02} corrected to the \loggf\ scale of the
NIST ASD \citep{kramida21}
}
\tablenotetext{d}{%
Abundance from \citet{cowan02} corrected to the \loggf\ scale of
\citet{biemont11}
}
\tablenotetext{e}{%
Abundance from \citet{cowan02} corrected to the \loggf\ scale of
\citet{ljung06}
}
\tablenotetext{f}{%
Mean abundance includes one line from \citet{roederer10b}
}
\tablenotetext{g}{%
Abundance from \citet{roederer10b} corrected to the \loggf\ scale
of \citet{hansen12}, including hyperfine splitting structure 
and isotope shifts
}
\tablenotetext{h}{%
See Appendix~\ref{appendix3}
}
\tablenotetext{i}{%
Mean abundance includes six lines from \citet{lawler07}
}
\tablenotetext{j}{%
Mean abundance includes one line from \citet{cowan05}
}
\tablenotetext{k}{%
Mean abundance includes three lines from \citet{cowan05}
}
\end{deluxetable}

\section{Tellurium and Platinum in HD~19445, HD~84937, and HD~140283}
\label{appendix2}

We use the STIS E230H spectrum of \hdone\
($R = 114,000$, 1879 $\leq \lambda \leq$ 2150~\AA,
GO-14672, datasets OD65A1010-A8030, PI Peterson) 
to derive its Te abundance and an upper limit on Pt.
We adopt the model atmosphere derived by 
\citet{peterson20},
\teff\ = 6070~K
\logg\ = 4.4,
[Fe/H] = $-2.15$, and
\vt\ = 1.3~\kmsec.
We derive abundances by spectrum synthesis matching
using the MOOG ``synth'' driver,
with line lists generated using LINEMAKE.~
All atomic data for these lines are
identical to the data adopted by \citet{roederer22a}.
We detect one Te~\textsc{i} line,
$\lambda$2142,
which yields \logeps{Te} = $0.65 \pm 0.15$,
or 
[Te/Fe] = $+0.62 \pm 0.15$.
No Pt~\textsc{i} lines are detected
in this spectrum.
One of the strongest ones, at 
$\lambda$2659, yields an upper limit
\logeps{Pt} $< +0.20$, 
or 
[Pt/Fe] $< +0.73$.

We use the STIS E230H spectrum of \hdeight\
($R = 114,000$, 2128 $\leq \lambda \leq$ 3143~\AA,
GO-14161, datasets OCTKA6010-D020, PI Peterson) 
to derive its Te abundance and an upper limit on Pt.
We adopt the model atmosphere derived by 
\citet{peterson20},
\teff\ = 6300~K,
\logg\ = 4.0,
[Fe/H] = $-2.25$, and
\vt\ = 1.3~\kmsec.
We detect two Te~\textsc{i} lines,
$\lambda$2142 and $\lambda$2385,
which yield \logeps{Te} abundances of
$0.29 \pm 0.10$ and $0.44 \pm 0.20$, respectively.
The mean Te abundance is
\logeps{Te} = $0.33 \pm 0.11$, or
[Te/Fe] = $+0.40 \pm 0.15$.
No Pt~\textsc{i} lines are detected
in this spectrum.
The Pt~\textsc{i} line at 2659~\AA\ 
yields an upper limit
\logeps{Pt} $< -0.10$,
or 
[Pt/Fe] $< +0.53$.

We use two STIS E230H spectra
($R = 114,000$, 1932 $\leq \lambda \leq$ 2212~\AA,
GO-7348, datasets O55Z01030-01050, PI Edvardsson;
$R = 50,000$, 2390 $\leq \lambda \leq$ 3140~\AA,
GO-9455, datasets O6LM71010-40, PI Peterson)
to derive upper limits on Te and Pt in \hdonefour.
We adopt the model atmosphere derived by
\citet{roederer12c},
\teff\ = 5600~K,
\logg\ = 3.66, 
[Fe/H] = $-2.62$, and
\vt\ = 1.15~\kmsec.
The Te~\textsc{i} line at 2142~\AA\
appears to be broadened by an Fe~\textsc{i} line,
so we use the Te~\textsc{i} line at 2259~\AA\
to derive \logeps{Te} $< 0.20$,
or [Te/Fe] $< +0.59$.
The Pt~\textsc{i} line at 2659~\AA\
yields an upper limit
\logeps{Pt} $< -1.40$, 
or
[Pt/Fe] $< -0.45$.

\section{Lutetium and Hafnium in HD~108317 and HD~128279}
\label{appendix3}

New atomic data motivate a reanalysis of the
lutetium (Lu, $Z = 71$) and hafnium (Hf, $Z = 72$)
abundances in 
\hdonezero\ and \hdonetwo.
\citet{denhartog20} measured the
hyperfine splitting constants for 16 levels of
ionized $^{175}$Lu,
the dominant Lu isotope,
and presented new line component patterns for 
35 Lu~\textsc{ii} transitions.
Furthermore, we identify
a moderately strong line at
the wavelength of the Lu~\textsc{ii} line at 
2615.41~\AA\
in several stars
with low levels of heavy \rpro\ elements
(see, e.g., the observational material listed in
table~3 of \citealt{roederer21}).
This line yields high Lu abundances,
\logeps{Lu/Eu} $\gtrsim 0.5$.
Such high ratios are not likely,
and they disagree with \logeps{Lu/Eu} ratios
($\approx -0.4$)
derived from other Lu~\textsc{ii} lines
in the highly \rpro-enhanced star \hdtwo\ \citep{roederer22a}.
We conclude that some of the
absorption at this wavelength is not due to Lu~\textsc{ii},
at least in stars without high levels of \rpro\ enhancement,
such as \hdtwo.
We revise the Lu abundances in \hdonezero\ and \hdonetwo\
to be upper limits:\
\logeps{Lu} $< -1.64$ and $< -1.92$, respectively.

\citet{denhartog21hf} 
measured the branching fractions for 199 transitions
of ionized Hf
and presented new \loggf\ values calculated
from these measurements.
We do not detect Hf~\textsc{ii} lines at 
2322.48, 2861.70, or 3012.90~\AA\
in either star.
The upper limits derived from these lines
are lower than the Hf abundances
derived previously from the Hf~\textsc{ii} lines at
2641.41 or 4093.15~\AA\ \citep{roederer12d},
so we conclude that the previous results
included unidentified blends.
We recommend 
\logeps{Hf} $< -1.40$
and 
\logeps{Hf} $< -1.70$ (both from the $\lambda$2322 line)
in \hdonezero\ and \hdonetwo, respectively.

\section{The Complete Heavy-Element Abundance Patterns}
\label{appendix4}

Table~\ref{fullabundtab} lists the
heavy-element abundance patterns
shown in Figure~\ref{abundplot}.
It also includes the stellar parameters adopted
by previous studies.
Any future study that makes use of these data
should cite the original references, which are
listed in Table~\ref{littab}.

%\startlongtable
\begin{deluxetable*}{cccccccccc}
\tablecaption{Stellar Parameters and Heavy-Element 
Abundance Patterns for the Sun and Eight Stars in the Sample
\label{fullabundtab}}
%\tablewidth{0pt}
\tabletypesize{\small}
\tablehead{
\colhead{Parameter} &
\colhead{Sun} & 
\colhead{HD 222925} &
\colhead{\bd} &
\colhead{HD 108317} &
\colhead{HD 160617} &
\colhead{HD 84937} &
\colhead{HD 19445} &
\colhead{HD 128279} &
\colhead{HD 140283} \\
\colhead{or abundance} &
\colhead{} &
\colhead{} &
\colhead{} &
\colhead{} &
\colhead{} &
\colhead{} &
\colhead{} &
\colhead{} &
\colhead{} 
}
\startdata
\teff\ (K)  & \nodata &  5636   &  5200   &  5100   &  5950   &  6300   &  6070   &  5080   &  5750   \\
\logg\      & \nodata &  2.54   &  1.80   &  2.67   &  3.90   &  4.00   &  4.40   &  2.57   &  3.70   \\
\vt\ (\kmsec)&\nodata & 2.20    &  1.90   &  1.50   &  1.30   &  1.30   &  1.30   &  1.60   &  1.40   \\
{[Fe/H]}    & \nodata & $-$1.46 & $-$2.10 & $-$2.37 & $-$1.77 & $-$2.25 & $-$2.15 & $-$2.46 & $-$2.57 \\
\hline        
\logeps{Se} & 3.34 &    2.62 & \nodata &    1.28 &    1.56 &    1.23 &    1.56 &    0.55 &    0.65 \\
\logeps{Sr} & 2.87 &    1.98 &    1.09 &    0.48 &    1.22 &    1.04 &    1.00 & $-$0.18 &    0.07 \\
\logeps{Y}  & 2.21 &    1.04 &    0.00 & $-$0.42 &    0.25 & $-$0.03 &    0.13 & $-$0.89 & $-$0.77 \\
\logeps{Zr} & 2.58 &    1.74 &    0.83 &    0.45 &    1.05 &    0.65 &    0.82 &    0.00 & $-$0.05 \\
\logeps{Nb} & 1.46 &    0.71 & $-$0.22 & $-$0.86 & \nodata & \nodata & \nodata & $-$0.88 & \nodata \\
\logeps{Mo} & 1.88 &    1.36 &    0.14 & $-$0.05 &    0.61 &    0.39 &    0.48 & $-$0.56 & $-$0.44 \\
\logeps{Ru} & 1.75 &    1.32 &    0.34 & $-$0.04 &    0.59 &    0.47 &    0.44 & $-$0.63 & \nodata \\
\logeps{Rh} & 0.91 &    0.64 & $-$0.57 & $-$0.84 & \nodata & \nodata & \nodata & \nodata & \nodata \\
\logeps{Pd} & 1.57 &    1.05 & $-$0.05 & $-$0.67 & $-$0.07 & \nodata & \nodata & $-$1.24 & \nodata \\
\logeps{Ag} & 0.94 &    0.44 & $-$0.71 & $-$1.34 & $-$0.89 & \nodata & \nodata & \nodata & \nodata \\
\logeps{Cd} & 1.71 &    0.34 & $-$0.99 & $-$1.15 & $-$0.03 & $-$0.17 & $-$0.37 & $-$1.32 & $-$1.34 \\
\logeps{In} & 0.80 &    0.51 &$<-$0.10 & \nodata & \nodata & \nodata & \nodata & \nodata & \nodata \\
\logeps{Sn} & 2.04 &    1.39 & \nodata & \nodata &    0.38 & $-$0.44 &    0.11 & \nodata & $-$1.06 \\
\logeps{Sb} & 1.01 &    0.37 & \nodata & \nodata & \nodata & \nodata & \nodata & \nodata & \nodata \\
\logeps{Te} & 2.18 &    1.63 &    0.42 &    0.06 &    0.67 &    0.33 &    0.65 & $-$0.12 & $<$0.20 \\
\logeps{Ba} & 2.18 &    1.26 &    0.48 & $-$0.32 &    0.44 & $-$0.23 &    0.00 & $-$1.03 & $-$1.09 \\
\logeps{La} & 1.10 &    0.51 & $-$0.42 & $-$1.12 & $-$0.36 & \nodata & \nodata & $-$1.64 & \nodata \\
\logeps{Ce} & 1.58 &    0.85 & $-$0.11 & $-$0.68 &    0.06 & \nodata & \nodata & $-$1.15 & \nodata \\
\logeps{Pr} & 0.72 &    0.22 & $-$0.71 & $-$0.90 & \nodata & \nodata & \nodata & \nodata & \nodata \\
\logeps{Nd} & 1.42 &    0.88 & $-$0.09 & $-$0.71 & $-$0.03 & \nodata & \nodata & $-$1.23 & \nodata \\
\logeps{Sm} & 0.96 &    0.62 & $-$0.34 & $-$1.05 & $-$0.45 & \nodata & \nodata & $-$1.50 & \nodata \\
\logeps{Eu} & 0.52 &    0.38 & $-$0.68 & $-$1.37 & $-$0.81 & $-$1.35 & $-$1.26 & $-$1.96 & $-$2.27 \\
\logeps{Gd} & 1.07 &    0.82 & $-$0.14 & $-$0.82 & $-$0.30 & \nodata & \nodata & \nodata & \nodata \\
\logeps{Tb} & 0.30 &    0.18 & $-$0.91 & $-$1.30 & \nodata & \nodata & \nodata & \nodata & \nodata \\
\logeps{Dy} & 1.10 &    1.01 & $-$0.04 & $-$0.75 & $-$0.20 & \nodata & \nodata & $-$1.35 & \nodata \\
\logeps{Ho} & 0.48 &    0.12 & $-$0.70 & $-$1.50 & \nodata & \nodata & \nodata & $-$1.71 & \nodata \\
\logeps{Er} & 0.92 &    0.73 & $-$0.25 & $-$0.94 & $-$0.35 & \nodata & \nodata & $-$1.54 & \nodata \\
\logeps{Tm} & 0.10 & $-$0.09 & $-$1.12 & $-$1.80 & \nodata & \nodata & \nodata & $-$2.24 & \nodata \\
\logeps{Yb} & 0.84 &    0.55 & $-$0.27 & $-$1.13 & $-$0.42 & \nodata & \nodata & $-$1.69 & \nodata \\
\logeps{Lu} & 0.10 & $-$0.04 & $-$1.33 &$<-$1.64 & \nodata & \nodata & \nodata &$<-$1.92 & \nodata \\
\logeps{Hf} & 0.85 &    0.32 & $-$0.66 &$<-$1.40 & \nodata & \nodata & \nodata &$<-$1.70 & \nodata \\
\logeps{W}  & 0.85 &    0.02 & \nodata & \nodata & \nodata & \nodata & \nodata & \nodata & \nodata \\
\logeps{Re} & 0.26 &    0.16 & \nodata & \nodata & \nodata & \nodata & \nodata & \nodata & \nodata \\
\logeps{Os} & 1.40 &    1.17 & $-$0.11 & $-$0.69 & $-$0.12 & \nodata & \nodata & \nodata & \nodata \\
\logeps{Ir} & 1.38 &    1.28 &    0.15 & \nodata & \nodata & \nodata & \nodata & \nodata & \nodata \\
\logeps{Pt} & 1.62 &    1.45 &    0.53 & $-$0.40 &    0.44 &$<-$0.10 & $<$0.20 & \nodata &$<-$1.40 \\
\logeps{Au} & 0.92 &    0.53 & $-$0.60 & $-$1.64 & \nodata & \nodata & \nodata & \nodata & \nodata \\
\enddata
\end{deluxetable*}

\bibliographystyle{aasjournal}
%bibliography{../../iuroederer}
\bibliography{ms.blg}

\end{document}